\documentclass[fleqn,10pt]{wlscirep}
\usepackage[utf8]{inputenc}
\usepackage[T1]{fontenc}
\usepackage{caption}
\usepackage{subcaption}
\usepackage{pgfplots}
\definecolor{hotmagenta}{rgb}{1.0, 0.11, 0.81}
	
\DeclareRobustCommand{\hwplotka}{\raisebox{2pt}{\tikz{\draw[blue,solid,line width=1.5pt](0,0)--(0.75mm,0);}}}
\DeclareRobustCommand{\hwplotkb}{\raisebox{2pt}{\tikz{\draw[hotmagenta,solid,line width=1.5pt](0,0)--(2mm,0);}}}
\DeclareRobustCommand{\hwplotkc}{\raisebox{2pt}{\tikz{\draw[hotmagenta,solid,line width=1.5pt](0,0)--(0.75mm,0);}}}
\DeclareRobustCommand{\hwplotkd}{\raisebox{2pt}{\tikz{\draw[red,solid,line width=1.5pt](0,0)--(5mm,0);}}}

\DeclareRobustCommand{\hwplotsasa}{\raisebox{2pt}{\tikz{\draw[black,solid,line width=1.5pt](0,0)--(5mm,0);}}}
\DeclareRobustCommand{\hwplotsasb}{\raisebox{2pt}{\tikz{\draw[red,solid,line width=1.5pt](0,0)--(2mm,0);}}}
\DeclareRobustCommand{\hwplotsasc}{\raisebox{2pt}{\tikz{\draw[blue,solid,line width=1.5pt](0,0)--(1.5mm,0);}}}
\DeclareRobustCommand{\hwplotsasd}{\raisebox{2pt}{\tikz{\draw[blue,solid,line width=1.5pt](0,0)--(0.75mm,0);}}}
\DeclareRobustCommand{\hwplotsase}{\raisebox{2pt}{\tikz{\draw[hotmagenta,solid,line width=1.5pt](0,0)--(0.75mm,0);}}}

\title{Nonlocal Models in the Analysis of Brain Neurodegenerative Protein Dynamics with Application to Alzheimer's Disease}

\author[1*]{Swadesh Pal}
\author[1,2]{Roderick Melnik}
\affil[1]{MS2Discovery Interdisciplinary Research Institute, Wilfrid Laurier University, Waterloo, Canada}
\affil[2]{BCAM- Basque Center for Applied Mathematics, Bilbao, Spain}

\affil[*]{spal@wlu.ca}


\begin{abstract}
It is well known that today nearly one in six of the world's population has to deal with neurodegenerative disorders. While a number of medical devices have been developed for the detection, prevention, and treatments of such disorders, some fundamentals of the progression of associated diseases are in urgent need of further clarification. In this paper, we focus on Alzheimer's disease, where it is believed that the concentration changes in amyloid-beta and tau proteins play a central role in its onset and development. A multiscale model is proposed to analyze the propagation of these concentrations in the brain connectome. In particular, we consider a modified heterodimer model for the protein-protein interactions. Higher toxic concentrations of amyloid-beta and tau proteins destroy the brain cell. We have studied these propagations for the primary and secondary and their mixed tauopathy. We model the damage of a brain cell by the nonlocal contributions of these toxic loads present in the brain cells. With the help of rigorous analysis, we check the stability behaviour of the stationary points corresponding to the homogeneous system. After integrating the brain connectome data into the developed model, we see that the spreading patterns of the toxic concentrations for the whole brain are the same, but their concentrations are different in different regions. Also, the time to propagate the damage in each region of the brain connectome is different. 
\end{abstract}
\begin{document}

\flushbottom
\maketitle
%
%
\thispagestyle{empty}

\section{Introduction}{\label{SE1}}

Alzheimer's disease (AD) is the most prevalent neurodegenerative disorder that slowly increases degrees of dementia over time \cite{C137152}. In most people with AD, the first symptoms appear after $60$ years of age, whereas some persons reported AD before 60 years of age, and it is not frequent. According to the Alzheimer's Association, five years ago $5.5$ million Americans in the U.S. and $44$ million worldwide dealt with this disease in all ranges of ages \cite{A325272}. At the same time, $10\%$ of the population over $65$ years of age is suffering, and it is predicted to be a public health crisis in the coming decades caused by this disease. Baumgart et al. estimated that to increase to 131 million people by 2050 \cite{B2015}. The history of the disease goes back well over $100$ years ago, when Dr. Alois Alzheimer in $1906$ noticed that a woman died with an unusual mental illness, and the symptoms were memory loss, language problems, etc. He examined her dead brain and identified many amyloid plaques and neurofibrillary tangles (NFTs) in the brain tissue \cite{A1906,HM1998,HG2003}. Today a wealth of medical and bioengineering tools exist for the detection, prevention, and treatments of such disorders. Researchers believe that the toxic changes in the brain destroy the communications between neurons \cite{A325272,LJ2011}.

The research involving amyloid-beta ($A\beta$) in AD has progressed more quickly than tau protein ($\tau P$). Amyloid-beta is a small piece of a larger protein amyloid precursor protein (APP) \cite{A325272}. It is still largely unknown how amyloid-beta accumulates in the brain and initiates AD. According to ``the amyloid hypothesis'', it is believed that flaws in the processes governing the production, accumulation, or disposal of amyloid-beta are the primary cause of AD \cite{A325272,DJ2016}. Recently, the Food and Drug Administration (FDA) approved the first drug ``Aducanumab'' for the treatment of some cases of Alzheimer's disease \cite{A2021}. This drug is a human antibody that targets the amyloid-beta protein and helps to reduce amyloid plaques associated with AD. Researchers continue to study whether this medication affects a person's rate of cognitive decline over time. In the AD-affected brain, $A\beta$ develops slowly and has many stages of building. Initially, $A\beta$ forms small clusters (oligomers), then a chain of oligomers (fibrils), and then ``mats'' of fibrils called beta-sheets. In the final stage, beta-sheets clump with other substances, form plaques. These amyloid plaques block the blood vessels and disrupt cell-to-cell communications (blood circulations) and active immune cells. These immune cells trigger inflammation and destroy the brain cells. Due to the failure of many clinical trials, researchers believe that $A\beta$ is not the only factor in the AD onset and progression. There must be some other factor involved in the disease.

There is a lot of evidence of the spread of tau protein with AD progressing \cite{RDF514524,C621,SL2018}. Along with the accumulation of $A\beta$ in the brain, scientists have also focussed on the importance of tau protein in AD. The tau proteins are neuronal microtubule (MT)-associated proteins found in the axon. Under physiological conditions, tau acts like a highly soluble and unfolded protein. It interacts with tubulin and promotes its assembly into MTs, which helps to stabilize their structure \cite{RDF514524,MAS1975}. Also, the tau protein plays an essential role in the balance between MT-dependent axonal transport of organelles and biomolecules \cite{GT2005,RJ2008}. At the first stage of AD progression, tau protein continuously accumulates in the entorhinal cortex and hippocampus regions. Then tau protein spreads into extracellular space in the form of oligomers \cite{P135}. Oligomers travel through a structure known as the synapse. It allows passing electrical or chemical signals from one brain cell to another through diffusion \cite{C621}. This abnormal tau distribution over the brain causes further disease progression. 

Many cellular mechanisms and their interactions between two different groups of proteins are not fully understood. In particular, the interactions between $A\beta$ and $\tau P$ are also, at a great degree unknown. Nowadays, researchers have focussed on finding the possible protein-protein interactions between two or more groups of proteins in AD \cite{M532540,SL2018,I20170777}. Following Walker et al., for AD, ``the amyloid-beta-tau nexus is central to disease-specific diagnosis, prevention and treatment'' \cite{L261265}. Thompson et al. observed some crucial interactions of these proteins \cite{TPEA1008267}: (i) amyloid-beta enhances the new seeding of toxic tau protein, (ii) toxic concentration of amyloid-beta depends on the concentration of tau protein, and (iii) amyloid-beta and tau protein enhance each other's toxicity.

A mathematical model is an excellent computational tool to work with these interactions. Many researchers considered these protein-protein interactions in mathematical models related to AD \cite{M2016,J2018,A4557,J2019,FSG2019a,FSG2020}. However, some questions regarding AD remain open, and more refined models need to be developed to address them. In this work, we apply a multiscale model that describes the interaction of two different protein families: $A\beta$ and $\tau P$. Each of the proteins follows the dynamics of a modified heterodimer model with local and nonlocal interactions and a coupling parameter between two proteins \cite{F109117,PC2014,TPEA1008267,SR2021,SR2021CP}.  Motivated by Meisl et al. \cite{GM2021}, we replace exponential growth with logistic growth in both the proteins $A\beta$ and $\tau P$. Due to toxic variants of these proteins present in the brain cell, neuronal damage occurs in the brain. We consider a model with local and nonlocal interactions for capturing such type of neuronal damage in the brain cells \cite{TPEA1008267}.

The rest of this manuscript is organized as follows. We consider a continuous nonlocal model for protein-protein interactions in Sect. \ref{SE2}. A nonlocal version of the damage of the cell is also described here. We have derived a nonlocal network model that can be used for integrating the brain connectome data. In Sect. \ref{SE3},  we discuss the temporal behaviour of the stationary points of the continuous mathematical model and provide details of the associated network model. The propagation of the toxic concentrations in the brain connectome, related to primary, secondary, and mixed tauopathies, is discussed in Sect. \ref{SE4}. Finally, the summary of the work is given in Sect. \ref{SE5}.

\section{Mathematical Model}{\label{SE2}} 

\bigskip

\subsection*{Continuous model}
We consider a spatial domain $\Omega$ in $\mathbb{R}^{3}$. For $x\in \Omega$ and time $t\in\mathbb{R}^{+}$, $u = u(\mathbf{x}, t)$ and $\widetilde{u} = \widetilde{u}(\mathbf{x}, t)$ represent the concentrations of healthy and toxic $A\beta$ respectively. Similarly, $v = v(\mathbf{x}, t)$ and $\widetilde{v} = \widetilde{v}(\mathbf{x}, t)$ represent the concentrations of healthy and toxic $\tau P$ respectively. Following the ideas of \cite{F109117,TPEA1008267,SR2021,SR2021CP}, the concentration evolution is given by the set of coupled integro-differential equations:
\begin{subequations}{\label{HCM1}}
\begin{align}
\frac{\partial u}{\partial t} & = \triangledown\cdot (\mathbf{D_{1}}\triangledown u) + a_{0} -a_{1}u - \frac{a_{2}u}{1+c_{u}u} \Phi\ast \widetilde{u}, \\
\frac{\partial \widetilde{u}}{\partial t} & =  \triangledown\cdot (\widetilde{\mathbf{D}}_{1}\triangledown \widetilde{u})  -\widetilde{a}_{1}\widetilde{u} + a_{2}\widetilde{u} \Phi\ast \bigg{(}\frac{u}{1+c_{u}u}\bigg{)},\\
\frac{\partial v}{\partial t} & = \triangledown\cdot (\mathbf{D_{2}}\triangledown v) + b_{0} -b_{1}v -b_{3}\widetilde{u}v\widetilde{v}-\frac{b_{2}v}{1+c_{v}v}\Phi\ast \widetilde{v},\\
\frac{\partial \widetilde{v}}{\partial t} & = \triangledown\cdot (\widetilde{\mathbf{D}}_{2}\triangledown \widetilde{v})  -\widetilde{b}_{1}\widetilde{v} +b_{3}\widetilde{u}v\widetilde{v}+b_{2}\widetilde{v}\Phi\ast \bigg{(}\frac{v}{1+c_{v}v}\bigg{)},
\end{align}
\end{subequations}
where the first two equations correspond to the healthy and toxic variants of the protein $A\beta$ and the last two equations play the same role for $\tau P$. Model (\ref{HCM1}) is an example of nonlocal models which become increasingly important in diverse areas of applications \cite{DR2021}. Here, $a_{0}$ and  $b_{0}$ are the mean production rates of healthy proteins, $a_{1}, b_{1}, \widetilde{a}_{1}$ and $\widetilde{b}_{1}$ are the mean clearance rates of healthy and toxic proteins, and $a_{2}$ and $b_{2}$ represent the mean conversion rates of healthy proteins to toxic proteins. The parameter $b_{3}$ is the coupling between the two proteins $A\beta$ and $\tau P$. $c_{u}$ and $c_{v}$ represent the conversion time of the healthy protein to the toxic protein of $A\beta$ and $\tau P$, respectively, and both are dimensionless parameters. The diffusion tensors $\mathbf{D}_{1}$, $\widetilde{\mathbf{D}}_{1}$, $\mathbf{D_{2}}$ and $\widetilde{\mathbf{D}}_{2}$ characterize the spreading of each proteins. For any time instance, the convolution term $\Phi \ast \widetilde{u}$ at the spatial point $\mathbf{x}$ is given by $$(\Phi \ast \widetilde{u})(\mathbf{x},t) =\int_{\Omega}\Phi (\mathbf{x-y})\widetilde{u}(\mathbf{y},t)d\mathbf{y}.$$ In a similar way, we define the other convolutions in (\ref{HCM1}). Here, $\Phi$ is the kernel function and it describes the conversion efficiencies between the spatial points $\mathbf{x}$ and $\mathbf{y}$. We assume that the kernel function is non-negative, even function, and has a compact support in $\mathbb{R}^{3}$. Also, $\Phi$ satisfies a standard normalized condition: $$\int_{\Omega}\Phi (\mathbf{x})d\mathbf{x} = 1.$$
Based on our application area, it is reasonable to assume that all variables and initial conditions are positive, and also that all the parameters are strictly positive.

Following Meisl et al.\cite{GM2021}, the fundamental model that includes the replication takes into account the logistic growth rather than exponential growth in the reaction-diffusion model. Therefore, we modify the exponential growth of healthy $A\beta$ and healthy $\tau P$ of the model (\ref{HCM1}) by the logistic growth that leads to the following system:
\begin{subequations}{\label{HCM2}}
\begin{align}
\frac{\partial u}{\partial t} & = \triangledown\cdot (\mathbf{D_{1}}\triangledown u) + u(a_{0} -a_{1}u) - \frac{a_{2}u}{1+c_{u}u} \Phi\ast \widetilde{u}, \\
\frac{\partial \widetilde{u}}{\partial t} & =  \triangledown\cdot (\widetilde{\mathbf{D}}_{1}\triangledown \widetilde{u})  -\widetilde{a}_{1}\widetilde{u} + a_{2}\widetilde{u} \Phi\ast \bigg{(}\frac{u}{1+c_{u}u}\bigg{)},\\
\frac{\partial v}{\partial t} & = \triangledown\cdot (\mathbf{D_{2}}\triangledown v) + v(b_{0} -b_{1}v) -b_{3}\widetilde{u}v\widetilde{v}-\frac{b_{2}v}{1+c_{v}v}\Phi\ast \widetilde{v},\\
\frac{\partial \widetilde{v}}{\partial t} & = \triangledown\cdot (\widetilde{\mathbf{D}}_{2}\triangledown \widetilde{v})  -\widetilde{b}_{1}\widetilde{v} +b_{3}\widetilde{u}v\widetilde{v}+b_{2}\widetilde{v}\Phi\ast \bigg{(}\frac{v}{1+c_{v}v}\bigg{)},
\end{align}
\end{subequations}
with appropriate initial and boundary conditions for all the components.

The above system (\ref{HCM2}) dictates the spread of two healthy and toxic variants of the proteins throughout the domain $\Omega$. The increase in the density of toxic proteins at a spatial point $\mathbf{x}$ in $\Omega$ disrupts the extracellular environment near $\mathbf{x}$ and causes intracellular activities. The impact of these toxic variants is not fully understood. However, Thompson et al. observed their correlation and defined a gross measure of regional neuronal damage by a function $q(\mathbf{x},t)\in [0,1]$ at a spatial point $\mathbf{x}$ at time $t$ \cite{TPEA1008267}. We say the neuron is healthy (functional) at a spatial point $\mathbf{x}$ if $q(\mathbf{x},t)=0$. On the other hand, if $q(\mathbf{x},t)=1$, then the neuron is no longer active, i.e., the neuron reached a fully damaged asymptotic state. The evolution of the damage is described as
\begin{equation}{\label{DE}}
\frac{dq}{dt}=(k_{1}\widetilde{u}+k_{2}\widetilde{v}+k_{3}\widetilde{u}\widetilde{v}+k_{4}\Psi \ast q)(1-q),
\end{equation}
with the initial condition $q(\mathbf{x},0)=0$, where $k_{1}$ and $k_{2}$ account for the damaging effect of toxic $A\beta$ and $\tau P$, respectively. The coefficient $k_{3}$ represents the damage due to the combined presence of both toxic loads. Finally, $k_{4}$ is the rate of transneuronal damage propagation, and it reflects aggregate neuronal damage from regional neighbours. Here, all the $k_{i}$'s are non-negative, and $\Psi \ast q$ denotes the convolution, defined as before. 

\subsection*{Network model}

Having defined the nonlocal continuous model, we are now in a position to develop a nonlocal network mathematical model that would account for the brain network data. The development of a coarse-grain mathematical model based on a continuous system is a key to defining a new network model for the brain connectome \cite{BM2019,FW2019}. Let $\mathbf{G}$ be a weighted brain graph with $N$ number of nodes and $E$ be the number of edges defined in the domain $\Omega$. Suppose, $\mathbf{W}$ is an adjacency matrix corresponding to the weighted graph $\mathbf{G}$. With the help of the adjacency matrix $\mathbf{W}$, we construct the Laplacian for the graph. For $i,j=1,2,3,\ldots,N$, we denote the elements of $\mathbf{W}$ as $$W_{ij}=\frac{n_{ij}}{l_{ij}^2},$$ where $n_{ij}$ is the mean fiber number and $l_{ij}^{2}$ is the mean length squared between the nodes $i$ and $j$. The Laplacian $\mathbf{L}$ of the graph $\mathbf{G}$ is a square matrix of order $N$ with the elements $$L_{ij}=\rho (D_{ii}-W_{ij}),~~i,j=1,2,3,\ldots,N,$$ where $\rho$ is the diffusion coefficient and $D_{ii}=\sum_{j=1}^{N}W_{ij}$ are the elements of the diagonal weighted degree matrix. 

Now, we begin to define the convolution term in the network model similar to the continuous model (\ref{HCM2}) for each of the nodes $j=1,2,3,\ldots,N$. In the continuous integro-differential model, the convolution term at a spatial point $\mathbf{x}$ is an integral in the neighbourhood (inside the domain of definition) of the spatial point $\mathbf{x}$ with Gaussian weight function \cite{SR2021CP}. Again, for the continuous model, all the neighbourhood points of $\mathbf{x}$ are connected. We apply the same connectedness technique for finding the convolution term at each node in the network model. 

First, following \cite{SR2021CP}, we fix a node $j$ and define a set of nodes $S_{j,1}$ containing all directly connected nodes to the node $j$. Then, we find another set of nodes $S_{j,2}$ containing all the immediate connected nodes, connected with the nodes in the set $S_{j,1}$. We continue this process in the whole graph $\mathbf{G}$ until we find the complete set of connected nodes (directly or indirectly). In this procedure, we obtain ``complete sets'' of nodes $S_{j,1}, S_{j,2},\ldots, S_{j,m_{j}}$ connected with the node $j$ and let $S_{j}$ be the union of these sets of nodes. Now, we sort all the nodes in $S_{j}$ according to the ``shortest distance'' from the node $j$. Suppose, $n_{j}$ is the total number of nodes connected with the node $j$ including the self node $j$ and $k_{1}, k_{2},\ldots,k_{n_{j}}$ are the labels of the sorted connected nodes starting with the self node $j$, i.e., $k_{1}=j$. We define a set of $n_{j}$ elements of weights as $$M_{j}'=\bigg{\{}1,e^{-\eta^{2}(s_{jk_{2}})^{2}}, e^{-\eta^{2}(s_{jk_{3}})^{2}},\ldots,e^{-\eta^{2}(s_{jk_{n_{j}}})^{2}} \bigg{\}}, $$
where $s_{jk_{i}}$ denotes the shortest distance from the node $k_{i}$ to the node $j$ along the edges.\\
We normalize $M_{j}'$ as $M_{j}=M_{j}'/|M_{j}'|$, where $$|M_{j}'|=\bigg{(}1+e^{-\eta^{2}(s_{jk_{2}})^{2}}+ e^{-\eta^{2}(s_{jk_{3}})^{2}}+\cdots+e^{-\eta^{2}(s_{jk_{n_{j}}})^{2}}  \bigg{)}.$$\\
Now, we define a row vector $V_{j}$ of $N$ number of elements with non-zero elements being $M_{j}(1),M_{j}(2),$ $\ldots,M_{j}(n_{j})$ at the indices $k_{1},k_{2},\ldots,k_{n_{j}}$ respectively. Finally, we are ready to define the convolution at the node $j$ in the graph $\mathbf{G}$ as 
\begin{equation}{\label{CTNNM}}
\Phi_{j}\ast \widetilde{u}_{j}=\sum_{n=1}^{N}V_{j}(n)\widetilde{u}_{n}.
\end{equation}
 If the node $j$ is not connected with any other nodes, then $\Phi_{j}\ast \widetilde{u}_{j}=\widetilde{u}_{j}$. Similarly, we define the other convolutions.

Taking all the factors (Laplacian and convolutions), we build a nonlocal network mathematical model on the brain graph $\mathbf{G}$. Let, $u_{j}, \widetilde{u}_{j}$ be the concentrations of healthy and toxic $A\beta$ and $v_{j},\widetilde{v}_{j}$ be the concentrations of healthy and toxic $\tau P$ at the node $j$. Then for $j=1,2,3,\ldots,N$, the network equations corresponding to the continuous model (\ref{HCM2}) is a system of first order differential equations and it is given by
\begin{subequations}{\label{HCMN}}
\begin{align}
\frac{du_{j}}{d t} & = -\sum_{k=1}^{N}L_{jk}u_{k}+ u_{j}(a_{0} -a_{1}u_{j}) - \frac{a_{2}u_{j}}{1+c_{u}u_{j}}\Phi_{j}\ast\widetilde{u_{j}}, \\
\frac{d\widetilde{u}_{j}}{d t} & =  -\sum_{k=1}^{N}L_{jk}\widetilde{u}_{k} -\widetilde{a}_{1}\widetilde{u}_{j} + a_{2}\widetilde{u}_{j}\Phi_{j}\ast \bigg{(} \frac{u_{j}}{1+c_{u}u_{j}}\bigg{)},\\
\frac{dv_{j}}{dt} & = -\sum_{k=1}^{N}L_{jk}v_{k} + v_{j}(b_{0} -b_{1}v_{j}) -b_{3}\widetilde{u}_{j}v_{j}\widetilde{v}_{j}- \frac{b_{2}v_{j}}{1+c_{v}v_{j}}\Phi_{j}\ast\widetilde{v}_{j},\\
\frac{d\widetilde{v}_{j}}{dt} & = -\sum_{k=1}^{N}L_{jk}\widetilde{v}_{k}  -\widetilde{b}_{1}\widetilde{v}_{j} +b_{3}\widetilde{u}_{j}v_{j}\widetilde{v}_{j} + b_{2}\widetilde{v}_{j}\Phi_{j}\ast \bigg{(} \frac{v_{j}}{1+c_{v}v_{j}}\bigg{)},
\end{align}
\end{subequations}
with non-negative initial conditions.

In a similar fashion, we define the network model for the nonlocal version of the damage equation (\ref{DE}). Let $q_{j}$ be the neuronal damage of cell located at the node $j$. Then for all the nodes $j=1,2,3,\ldots,N$, the network damage equation corresponding to the network model is given by 
\begin{equation}{\label{NDM}}
\frac{dq_{j}}{dt}=(k_{1}\widetilde{u}_{j}+k_{2}\widetilde{v}_{j}+k_{3}\widetilde{u}_{j}\widetilde{v}_{j}+k_{4}\Psi \ast q_{j})(1-q_{j}),
\end{equation} 
with the initial condition $q_{j}=0$. Similar to the nonlocal interaction defined in (\ref{CTNNM}), we use the same technique to find the convolution $\Psi \ast q_{j}$ for the neuronal damage corresponding to the node $j$, but with a different controlling parameter $\sigma$ in the place of $\eta$.

\section{Analysis of the Continuous Model}{\label{SE3}}

Before analyzing the full reaction-diffusion model consisting of a coupled set of differential equations (\ref{HCM2}), we first study the temporal dynamics (e.g., equilibrium points and their stabilities) of the model. Temporal dynamics is generally a spatially independent dynamics of a reaction-diffusion model. Due to the spatially independent behaviours, the temporal dynamics for the local and nonlocal models are the same. Therefore, under these simplifications, the temporal model corresponding to the nonlocal model (\ref{HCM2}) is given by
\begin{subequations}{\label{HCMT}}
\begin{align}
\frac{du}{dt} & =  u(a_{0} -a_{1}u) - \frac{a_{2}u}{1+c_{u}u}\widetilde{u}, \\
\frac{d\widetilde{u}}{dt} & =   -\widetilde{a}_{1}\widetilde{u} + \frac{a_{2}u}{1+c_{u}u}\widetilde{u},\\
\frac{dv}{dt} & =  v(b_{0} -b_{1}v) -b_{3}\widetilde{u}v\widetilde{v} - \frac{b_{2}v}{1+c_{v}v}\widetilde{v},\\
\frac{d\widetilde{v}}{dt} & =  -\widetilde{b}_{1}\widetilde{v} +b_{3}\widetilde{u}v\widetilde{v} + \frac{b_{2}v}{1+c_{v}v}\widetilde{v},
\end{align}
\end{subequations}
with non-negative initial conditions. The system (\ref{HCMT}) has some feature points, such that characterize its behaviour but do not change over time, called equilibrium points. Note that these temporal dynamics apply to both continuous and network models. The stationary states or the solutions of the continuous and network models are the equilibrium points of the system (\ref{HCMT}).

The system (\ref{HCMT}) always has a trivial equilibrium point $E_{0} = (0, 0, 0, 0)$. This equilibrium point corresponds to the the absence of all the ingredients in the system. Generally, it does not occur in the living brain but occurs only in the dead brain cell. Depending on the parameter values, we have a maximum of seven equilibrium points of the system (\ref{HCMT}) where at least one component of the equilibrium point is zero. These are called the axial equilibrium points. The possible axial equilibrium points are
\begin{align*}
&E_{1} = \bigg{(}\frac{a_{0}}{a_{1}}, 0, 0, 0\bigg{)},\quad E_{2} = \bigg{(}0, 0, \frac{b_{0}}{b_{1}}, 0\bigg{)}, \quad E_{3} = \bigg{(}\frac{a_{0}}{a_{1}}, 0, \frac{b_{0}}{b_{1}}, 0\bigg{)},\\
&E_{4} = \bigg{(}\frac{\widetilde{a}_{1}}{a_{2}-c_{u}\widetilde{a}_{1}}, \frac{a_{0}(a_{2}-c_{u}\widetilde{a}_{1})-a_{1}\widetilde{a}_{1}}{(a_{2}-c_{u}\widetilde{a}_{1})^{2}}, 0, 0\bigg{)}, \quad E_{5} = \bigg{(}0,0,\frac{\widetilde{b}_{1}}{b_{2}-c_{v}\widetilde{b}_{1}}, \frac{b_{0}(b_{2}-c_{v}\widetilde{b}_{1})-b_{1}\widetilde{b}_{1}}{(b_{2}-c_{v}\widetilde{b}_{1})^{2}}\bigg{)} \\
& E_{6} = \bigg{(}\frac{\widetilde{a}_{1}}{a_{2}-c_{u}\widetilde{a}_{1}}, \frac{a_{0}(a_{2}-c_{u}\widetilde{a}_{1})-a_{1}\widetilde{a}_{1}}{(a_{2}-c_{u}\widetilde{a}_{1})^{2}}, \frac{b_{0}}{b_{1}}, 0\bigg{)}, \quad  E_{7} =  \bigg{(}\frac{a_{0}}{a_{1}}, 0, \frac{\widetilde{b}_{1}}{b_{2}c_{v}\widetilde{b}_{1}}, \frac{b_{0}(b_{2}-c_{v}\widetilde{b}_{1})-b_{1}\widetilde{b}_{1}}{(b_{2}-c_{v}\widetilde{b}_{1})^{2}}\bigg{)}.
\end{align*}
The concentration in each of the components cannot be negative. Therefore, for the existence of the equilibrium points $E_{4}$ and $E_{7}$, we must have $a_{2}>c_{u}\widetilde{a}_{1}$ and $a_{0}/a_{1}\geq \widetilde{a}_{1}/(a_{2}-c_{u}\widetilde{a}_{1})$. Similarly, for the feasibility of the equilibrium points $E_{5}$ and $E_{6}$, we need $b_{2}>c_{v}\widetilde{b}_{1}$ and $b_{0}/b_{1}\geq \widetilde{b}_{1}/(b_{2}-c_{v}\widetilde{b}_{1})$.

We call a stationary state healthy $A\beta$ (healthy $\tau P$) if the second (fourth) component of the equilibrium point is zero; otherwise, it is toxic $A\beta$ (toxic $\tau P$). The equilibrium points $E_{1}$ and $E_{2}$ contain either healthy $A\beta$ or healthy $\tau P$ concentrations, and these are healthy $A\beta$ and healthy $\tau P$ stationary states, respectively. $E_{3}$ is a ``healthy $A\beta$ - healthy $\tau P$'' stationary state since it does not have any of the toxic loads $A\beta$ or $\tau P$, i.e., no amyloid plaques or neurofibrillary tau tangles. Also, we call this equilibrium point the disease-free stationary state. The other equilibrium points are brain disease states since these brain states have either amyloid plaques or neurofibrillary tau tangles. Now, we are interested in finding an equilibrium point where both toxic loads are present. 

Suppose, $E_{*} = (u_{*},\widetilde{u}_{*},v,\widetilde{v}_{*})$ is a positive equilibrium point. Then we have $u_{*} = \widetilde{a}_{1}/(a_{2}-c_{u}\widetilde{a}_{1})$, $\widetilde{u}_{*} = (a_{0}a_{2}-a_{1}\widetilde{a}_{1}-c_{u}a_{0}\widetilde{a}_{1})/(a_{2}-c_{u}\widetilde{a}_{1})^{2}$, $\widetilde{v}_{*} = (b_{0}-b_{1}v_{*})(c_{v}v_{*}+1)/(b_{3}c_{v}\widetilde{u}_{*}v_{*}+b_{3}\widetilde{u}+b_{2})$, where $v_{*}$ satisfies the quadratic equation
\begin{equation}{\label{QEV}}
b_{3}c_{v}\widetilde{u}_{*}v_{*}^{2}+(b_{2}+b_{3}\widetilde{u}_{*}-\widetilde{b}_{1}c_{v})v_{*}-\widetilde{b}_{1} = 0.
\end{equation}
For the feasibility of the first two components of the equilibrium point $E_{*}$, we must have $a_{2}>c_{u}\widetilde{a}_{1}$ and $a_{0}/a_{1}\ > \widetilde{a}_{1}/(a_{2}-c_{u}\widetilde{a}_{1})$. Along with these conditions, we also require $b_{0}>b_{1}v_{*}$, where $v_{*}$ is a positive root of the equation (\ref{QEV}). In this case, the equilibrium point $E_{*}$ becomes ``toxic $A\beta$ - toxic $\tau P$'' stationary state as it contains both amyloid plaques and neurofibrillary tau tangles.

Now, we briefly discuss the stability of the equilibrium points of the system (\ref{HCMT}). The stability behaviour of an equilibrium point of the system depends only on the eigenvalues of the Jacobian matrix calculated at that point. For a fixed equilibrium point, if the real parts of all the eigenvalues of the Jacobian matrix are negative, the equilibrium point is stable; otherwise, it is unstable.  The Jacobian matrix of the system (\ref{HCMT}) about any equilibrium point $(u_{s},\widetilde{u}_{s},v_{s},\widetilde{v}_{s})$ is given by 
\begin{equation*}
\mathbf{J}_{s}=\begin{bmatrix}
J_{11} & J_{12} & 0 & 0 \\
J_{21} & J_{22} & 0 & 0 \\
0 & J_{32} & J_{33} & J_{34}\\
0 & J_{42} & J_{43} & J_{44}
\end{bmatrix},
\end{equation*} 
where $J_{11} = a_{0}-2a_{1}u_{s}-a_{2}\widetilde{u}_{s}/(1+c_{u}u_{s})^{2},$ $J_{12} = -a_{2}u_{s}/(1+c_{u}u_{s})$, $J_{21} = a_{2}\widetilde{u}_{s}/(1+c_{u}u_{s})^{2}$ $J_{22} = -\widetilde{a}_{1} + a_{2}u_{s}/(1+c_{u}u_{s}),$ $J_{32} = -b_{3}v_{s}\widetilde{v}_{s} ,$ $J_{33} = b_{0}-2b_{1}v_{s}-b_{3}\widetilde{u}_{s}\widetilde{v}_{s}-b_{2}\widetilde{v}_{s}/(1+c_{v}v_{s})^{2},$ $J_{34} = -b_{2}v_{s}/(1+c_{v}v_{s})-b_{3}\widetilde{u}_{s}v_{s},$ $J_{42} = b_{3}v_{s}\widetilde{v}_{s},$ $J_{43} = b_{3}\widetilde{u}_{s}\widetilde{v}_{s}+b_{2}\widetilde{v}_{s}/(1+c_{v}v_{s})^{2}$ and $J_{44} = -\widetilde{b}_{1}+b_{3}\widetilde{u}_{s}v_{s} + b_{2}v_{s}/(1+c_{v}v_{s})$.

Therefore, the eigenvalues of the Jacobian matrix $\textbf{J}_{s}$ are $\lambda_{1,2} = -(T\pm\sqrt{T^{2}- 4D})/2, \lambda_{3,4} = -(\widehat{T}\pm \sqrt{{\widehat{T}}^{2}- 4\widehat{D}})/2$, where $T =- (J_{11}+J_{22}), D = J_{11}J_{22}-J_{12}J_{21}, \widehat{T}  = -(J_{33}+J_{44})$ and $\widehat{D} = J_{33}J_{44} - J_{34}J_{43}$. Depending on the parameter values, the number of stationary points is different. Hence, in the next section, we discuss the stability of the stationary points after fixing the parameter values, motivated by \cite{TPEA1008267}.

\section{Results and Discussion}{\label{SE4}}

Tau is a microtubule-associated protein predominantly expressed in nerve cells promoting microtubule assembly and stabilization. It is a cytosolic protein mainly present in axons and involved in anterograde axonal transport. Tau protein alters its metabolism in several neurodegenerative diseases (e.g., AD). Thus alterations in the amount of the tau protein, missense mutations, post-transcriptional modifications, aberrant tau aggregation, or a different expression of some of its isoforms could provoke pathological effects resulting in the appearance of neuronal disorders known as tauopathies. Researchers are making a substantial effort to generate tau oligomers by purified recombinant protein strategies for studying oligomer conformations and their toxicity. Still, there is no specific toxic tau protein has been identified \cite{CH2021}. However, some cellular biosensor technology has been discovered to monitor the formation of tau oligomers and aggregates in the living cells \cite{CH2021}, e.g., fluorescence resonance energy transfer (FRET), bimolecular fluorescence complementation (BiFC), and split luciferase complementation (SLC).

Based on the pathology, there are two groups of disease propagation which are classified as: primary, when tau is the main lesion, and secondary, when tau is associated with other pathology \cite{CS2021,D2020}. Here, for the primary tauopathy, the non-zero concentration of toxic $\tau P$ exists independent of the concentration of $A\beta$. On the other hand, this dependency needs for the secondary tauopathy. After observing several characteristics of the models, we move to an interesting scenario in the brain network model where some regions in the brain satisfy the primary tauopathy, and the other regions satisfy the secondary tauopathy. This is called a mixed tauopathy, and it is more realistic than primary or secondary \cite{JD2021}. We also study the damage dynamics in each of the tauopathies. 

In this work, we use actual brain connectome data \cite{CBBV2017, CBBV2016,BCBV2019} for the network model available on the website https://braingraph.org. We choose high-resolution data consisting of 1,015 nodes and a different number of edges. We compared different data sets in the brain connectome to check the numerical artifact, and the results are consistent. We have used the C programming language and Matlab for all the computations and simulations.

\begin{figure}[]
\begin{center}
                \centering
                \includegraphics[width=10cm]{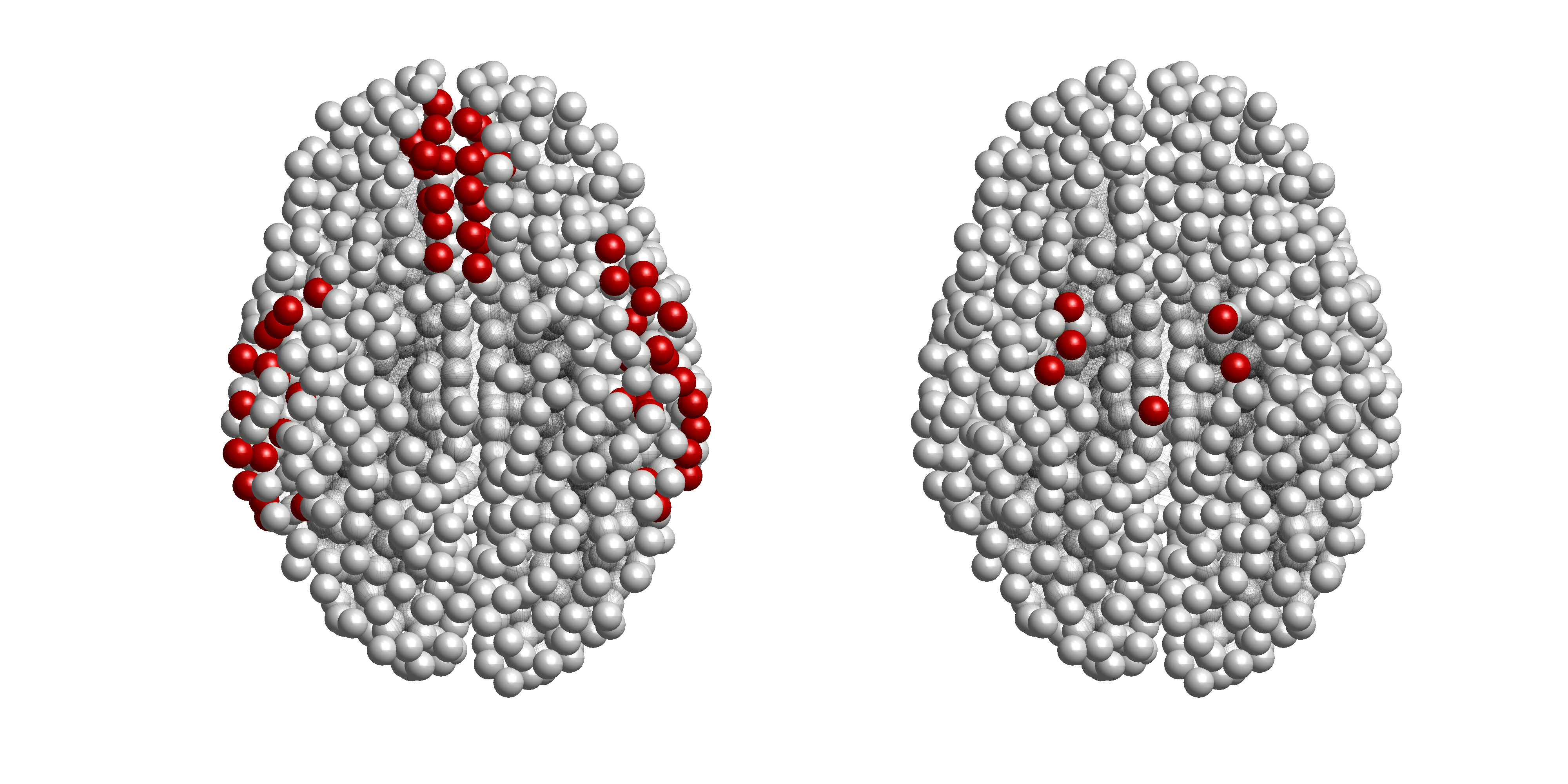}
        \caption{(Color online) Initial seeding sites for Alzheimer's disease: (left) toxic $A\beta$ and (right) toxic $\tau P$.}\label{fig:ubvbinitial}
\end{center}
\end{figure}

As the first step in the disease propagation analysis, we set seeding sites of toxic amyloid-beta and toxic tau proteins in the brain region(s) associated with Alzheimer's disease. In the AD-affected brain, researchers observed that the amyloid-beta plaques start their propagation from temporobasal and frontomedial regions in the brain \cite{TPEA1008267,M4551,M20312038,M532540,FSG2019a}. All the nodes in these two regions are the initial seeding sites for the toxic $A\beta$ (see Fig. \ref{fig:ubvbinitial}). On the other hand, locus coeruleus and transentorhinal associated regions are the initial tau staging sites for the AD \cite{GM2021,H239259} (see Fig. \ref{fig:ubvbinitial}). Seeding sites are highlighted in red, whereas the gray color corresponds to the zero toxic loads in the brain connectome. Initially, we set all the nodes to a healthy state (i.e., $E_{3}$) except the seeding nodes. We add small toxic loads $0.25\%$ and $0.38\%$ in the toxic $A\beta$ and toxic $\tau P$ concentrations for the seeding sites, respectively. In modelling other scenarios, one may choose some other percentages of the toxic loads for the computations. We use the same initial conditions every time for finding the solution of the network model.

\subsection{Primary and secondary tauopathies}

For the analysis of disease progression in the brain initially, we take a globally-constant synthetic parameter values throughout all the regions in the brain connectome. The first example is corresponding to the primary tauopathy. All the uniform parameter values for the primary tauopathy are in Table \ref{table:TFPV} except the coupling parameter $b_{3}$ and conversion times $c_{u}$ and $c_{v}$. The coupling parameter is not a passive facet of disease phenomenology but plays a much more integral role in secondary tauopathy \cite{TPEA1008267}. In the primary tauopathy, first we fix $b_{3} = 4.14$ and study the effect of conversion times $c_{u}$ and $c_{v}$ on the local network model.

\begin{table}[h!]
\caption{Fixed parameter values for the primary tauopathy}
\label{table:TFPV}
\begin{center}
\begin{tabular}{|c|c|c|c|}
\hline
Healthy $A\beta$ & Toxic $A\beta$ & Healthy $\tau P$ & Toxic $\tau P$ \\
\hline
$\rho = 1$ & $ \rho = 1$ & $\rho= 1$ & $\rho= 1$  \\
\hline
$a_{0} = 1.035$ & $\widetilde{a}_{1} = 0.828$ & $b_{0}= 0.69$ & $\widetilde{b}_{1}= 0.552$  \\
\hline
$a_{1} = 1.38$ & $a_{2} = 1.38$ & $b_{1}= 1.38$ & $b_{2}= 1.38$\\
\hline
\end{tabular}
\end{center}
\end{table}

\begin{figure}[]
\begin{center}
     \centering
        \begin{subfigure}[p]{0.31\textwidth}
                \centering
                \includegraphics[width=\textwidth]{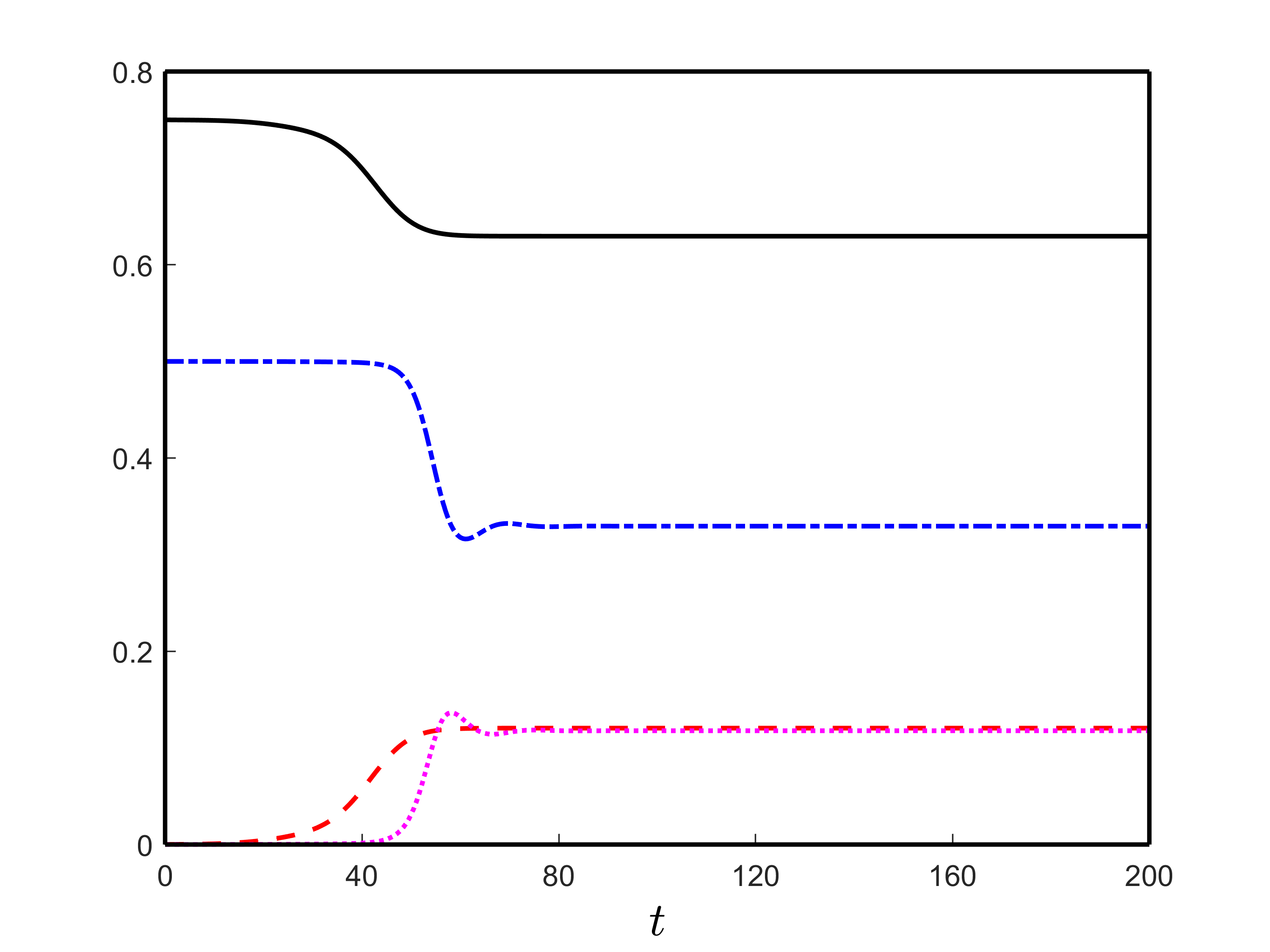}
                \caption{ }\label{fig:SASLME1a}
        \end{subfigure}
        ~
        \begin{subfigure}[p]{0.31\textwidth}
                \centering
                \includegraphics[width=\textwidth]{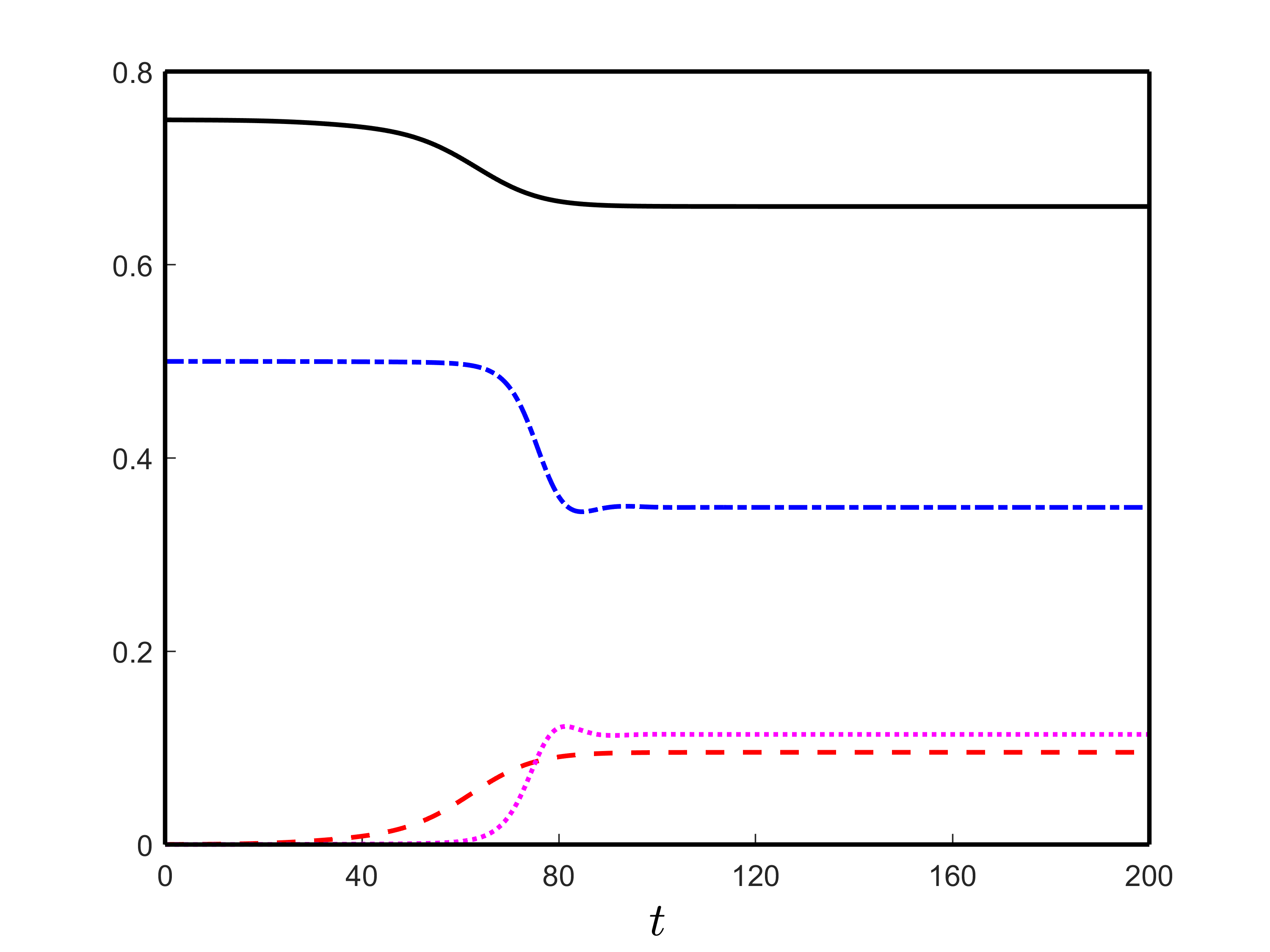}
                \caption{ }\label{fig:SASLME1b}
        \end{subfigure}
        ~
        \begin{subfigure}[p]{0.31\textwidth}
                \centering
                \includegraphics[width=\textwidth]{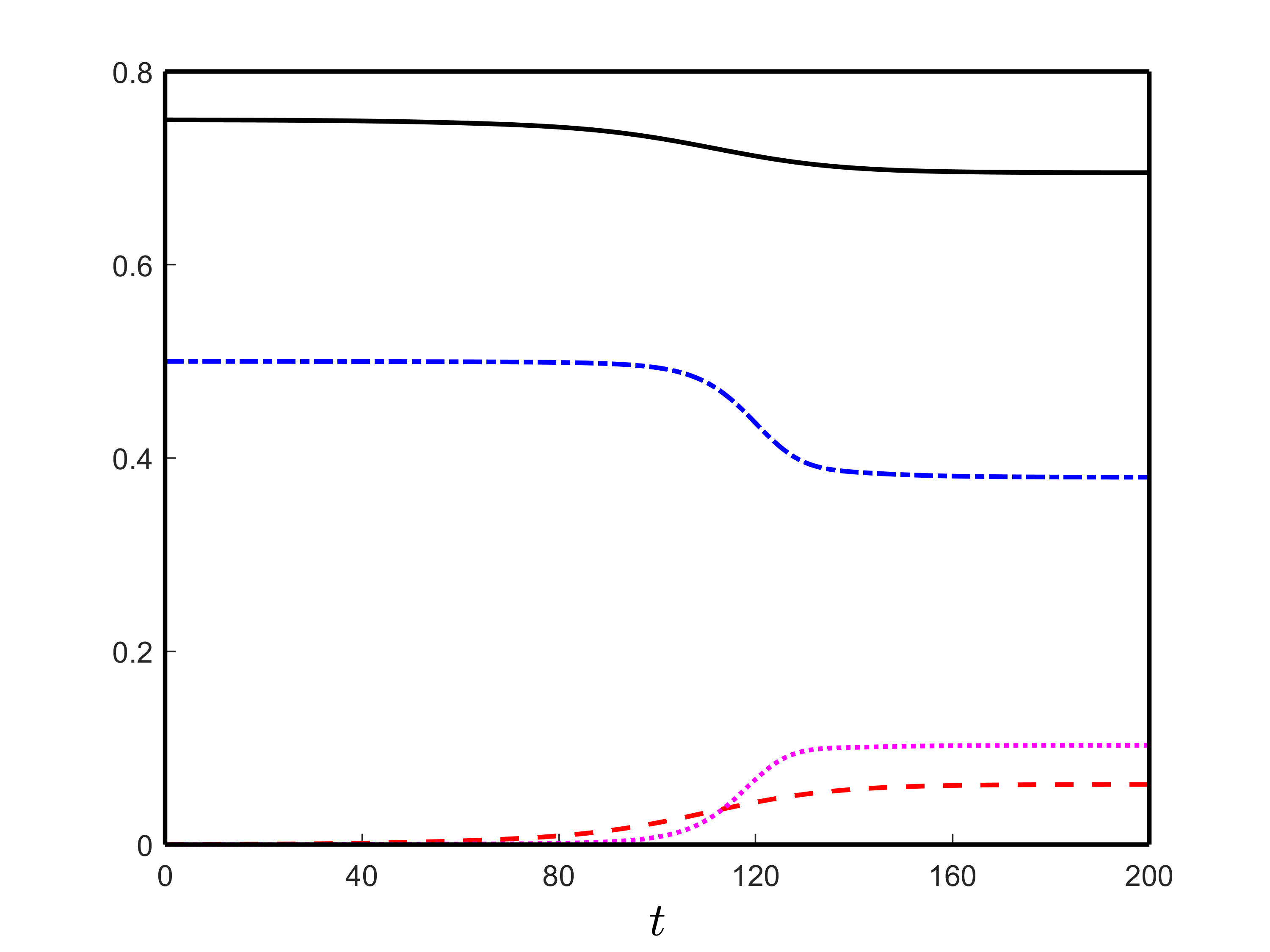}
                \caption{ }\label{fig:SASLME1c}
        \end{subfigure}
        \caption{ (Color online) Spatial averages of the solutions of the local model with different values of $c_{u}$ and $c_{v}$: (\subref{fig:SASLME1a}) $c_{u}=c_{v} = 0$, (\subref{fig:SASLME1b}) $c_{u}=c_{v} = 0.1$ and (\subref{fig:SASLME1c}) $c_{u}=c_{v} = 0.2$. The curves (\hwplotsasa), (\hwplotsasb \hwplotsasb), (\hwplotsasc \hwplotsasd \hwplotsasc) and (\hwplotsase \hwplotsase \hwplotsase \hwplotsase) are corresponding to the spatial averages of $u$, $\widetilde{u}$, $v$ and $\widetilde{v}$, respectively.}\label{fig:SASLME1}
\end{center}
\end{figure}

Firstly, we take $c_{u}=c_{v}=0$. All possible stationary states exist (described in the earlier section) for the system with these parameter values. Here, all the stationary states are unstable except for the positive stationary state $E_{*}$. Therefore, the disease propagates from the initial seeding sites and converges to the positive steady state $E_{*}$ of the local system. We plot the spatial averages of the solutions of the local network model in Fig. \ref{fig:SASLME1}(\subref{fig:SASLME1a}). Similarly, the spatial averages of the solutions of the local network model for two other values of the conversion times are shown in Figs. \ref{fig:SASLME1} (\subref{fig:SASLME1b}) and (\subref{fig:SASLME1c}). With an increase in the conversion time, the required time to propagate the disease in the whole brain connectome increases.

\begin{figure}[]
\begin{center}
                \centering
                \includegraphics[width=17cm]{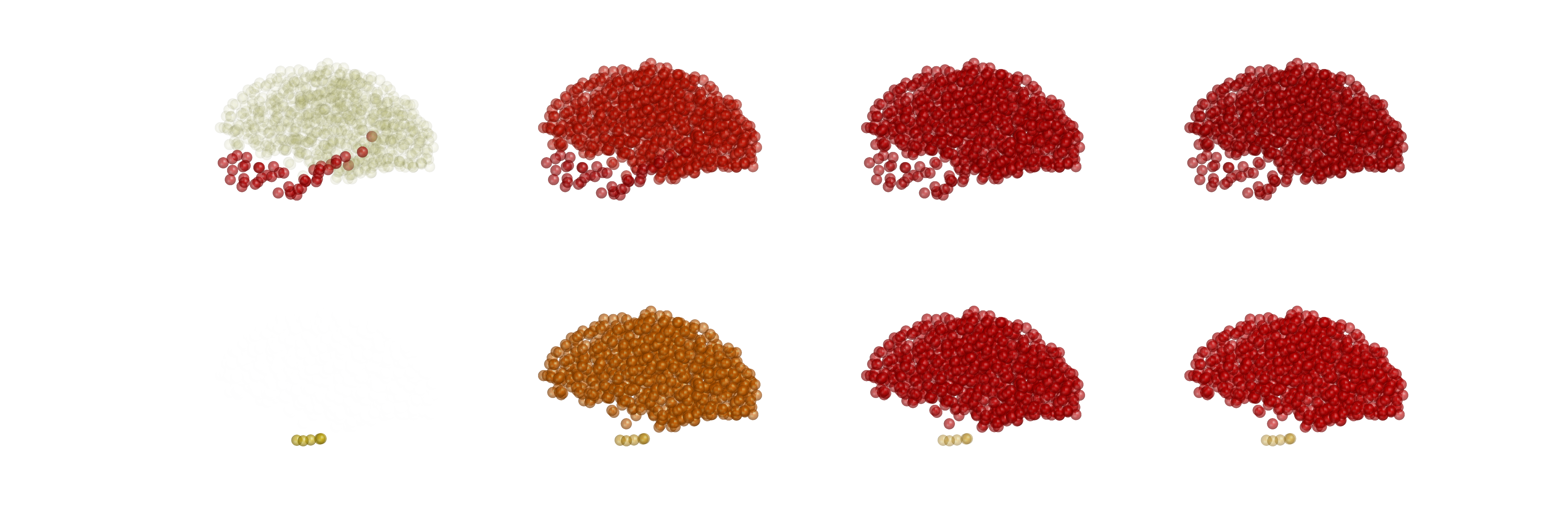}
        \caption{(Color online) Toxic $A\beta$ and toxic $\tau P$ concentrations in the brain connectome at different times (left to right: $t=50, 75, 85$ and $100$) for the local model corresponding to the primary tauopathy. Upper and lower panels are corresponding to toxic $A\beta$ and toxic $\tau P$, respectively.}\label{fig:ubvbpt}
\end{center}
\end{figure}

We see some spikes in the spatial-average solution of toxic $\tau P$ in Fig. \ref{fig:SASLME1}. Here, two eigenvalues of the Jacobian matrix around the positive stationary states are negative real numbers, but the other two are complex numbers with negative real parts. Therefore, the solution converges to the stable positive stationary state, but oscillations happen in a small neighbourhood of the point before converging to the solution. With an increase in the conversion time, the amplitude of the oscillations decrease (see Figs. \ref{fig:SASLME1} (\subref{fig:SASLME1b}) and (\subref{fig:SASLME1c})). Generally, at the initial stage, toxic $\tau P$ grows at the seeding sites and distributes the concentration throughout the brain connectome. After this spreading stage, toxic $\tau P$ concentration rises in the whole brain. So the average density of toxic $\tau P$ grows very fast, and it crosses the stationary state component corresponding to toxic $\tau P$. Generally, the cerebrovascular system reduces the toxic density in the brain and keeps it free from harmful agents. Due to the high accumulation rate of toxic $\tau P$ all over the brain, the cerebrovascular system can not control the whole toxic density. At last, it reduces some toxic levels of $\tau P$, not the total concentrations, and saturates to the disease state ``toxic $A\beta$ - toxic $\tau P$''.

\begin{figure}[ht!]
\begin{center}
                \centering
                \begin{subfigure}[p]{0.4\textwidth}
                \centering
                \includegraphics[width=\textwidth]{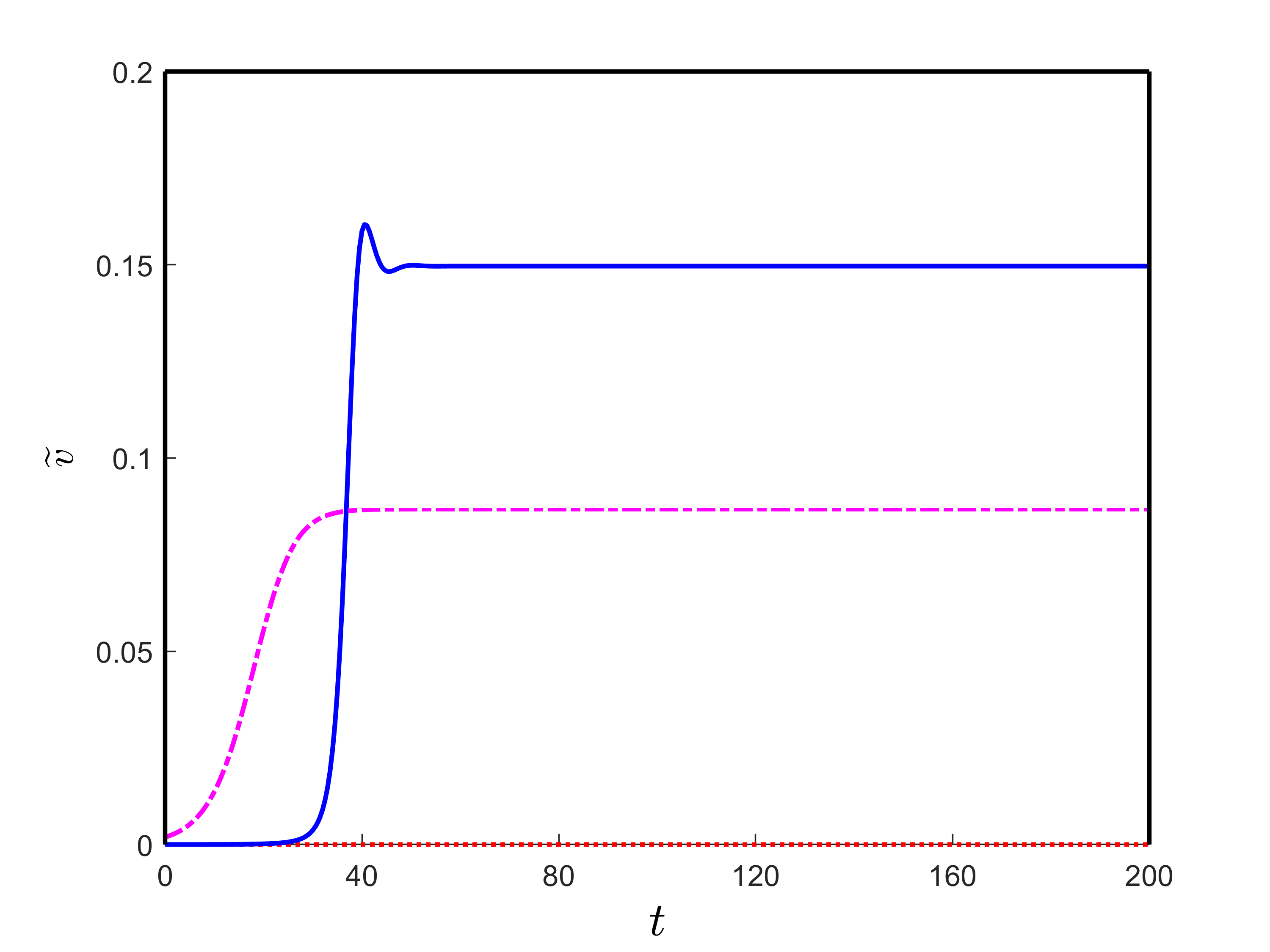}
                \caption{ }\label{fig:rwvbsa}
        \end{subfigure}
        ~
        \begin{subfigure}[p]{0.4\textwidth}
                \centering
                \includegraphics[width=\textwidth]{local_e1_0_1_tvb.png}
                \caption{ }\label{fig:rwvbsb}
        \end{subfigure}
        \caption{(Color online) Node-wise toxic propagations over time: (\subref{fig:rwvbsa}) toxic $A\beta$ and (\subref{fig:rwvbsa}) toxic $\tau P$.}\label{fig:rwvbs}
\end{center}
\end{figure}

Now, we shall analyze the general dynamics at each node located in different regions in the brain connectome. We provide the parameter values in Table \ref{table:TFPV} along with $b_{3} = 4.14$ and $c_{u} = c_{v} = 0.1$. The distributions of toxic $A\beta$ and toxic $\tau P$ in the brain connectome are shown in Fig. \ref{fig:ubvbpt}. Here, non-uniform density distributions of all four components exist in the brain connectome, and they occur due to the initial seeding of toxic loads in some particular regions. To verify the non-uniformity, we plot the solution profiles for toxic $A\beta$ and toxic $\tau P$ corresponding to each of the nodes in Fig. \ref{fig:rwvbs}. Here, we see three different solution behaviours for toxic $A\beta$ and $\tau P$. For the case of toxic $A\beta$, the dashed-dotted, solid, and dotted curves correspond to the seeding connected nodes (temporobasal and frontomedial regions), disconnected nodes, and the rest of the nodes, respectively. The same types of curves are plotted for toxic $\tau P$. 

\begin{figure}[ht!]
\begin{center}
                \centering
                \begin{subfigure}[p]{0.4\textwidth}
                \centering
                \includegraphics[width=\textwidth]{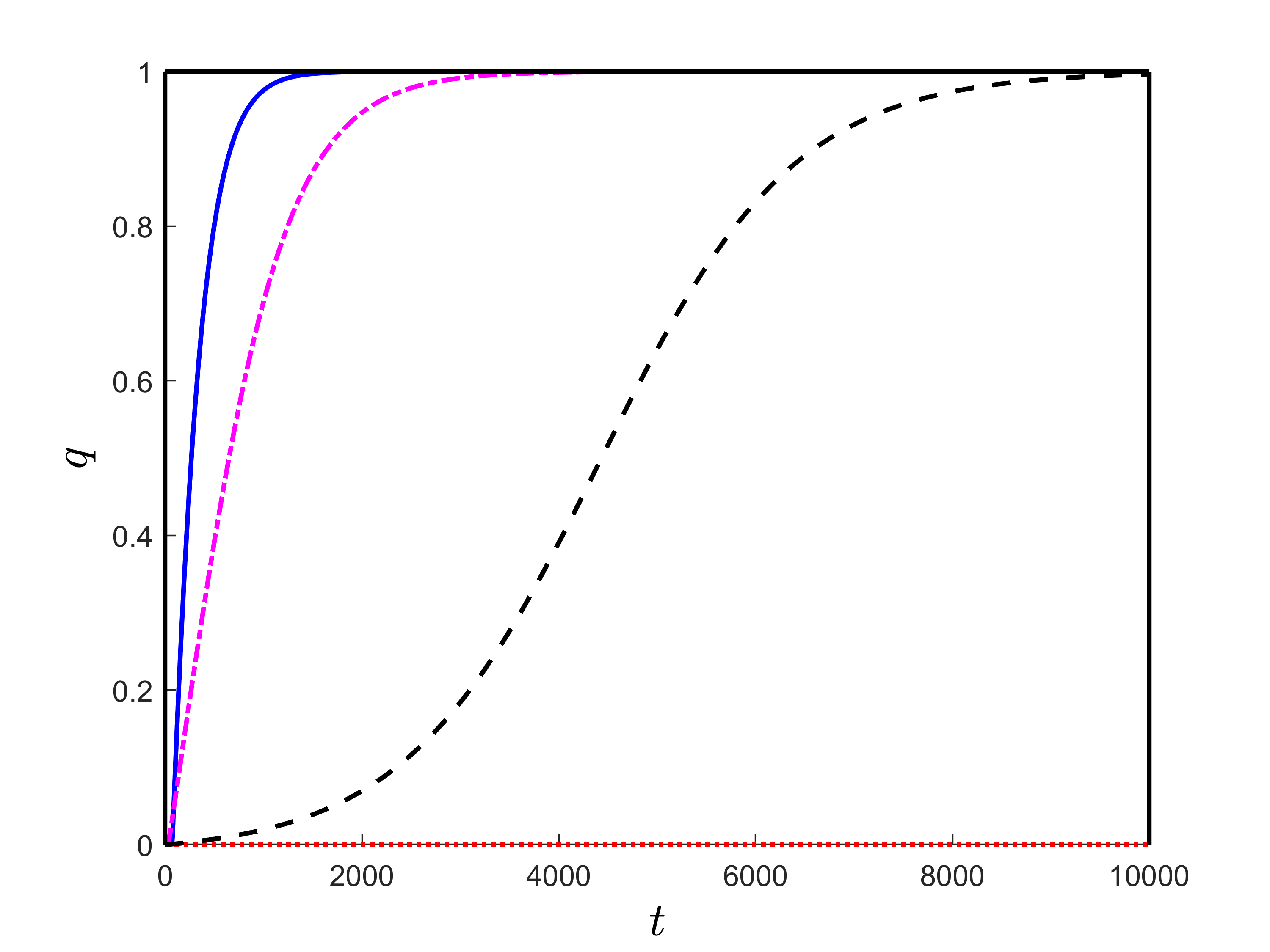}
                \caption{ }\label{fig:RWDSa}
        \end{subfigure}
        ~
        \begin{subfigure}[p]{0.4\textwidth}
                \centering
                \includegraphics[width=\textwidth]{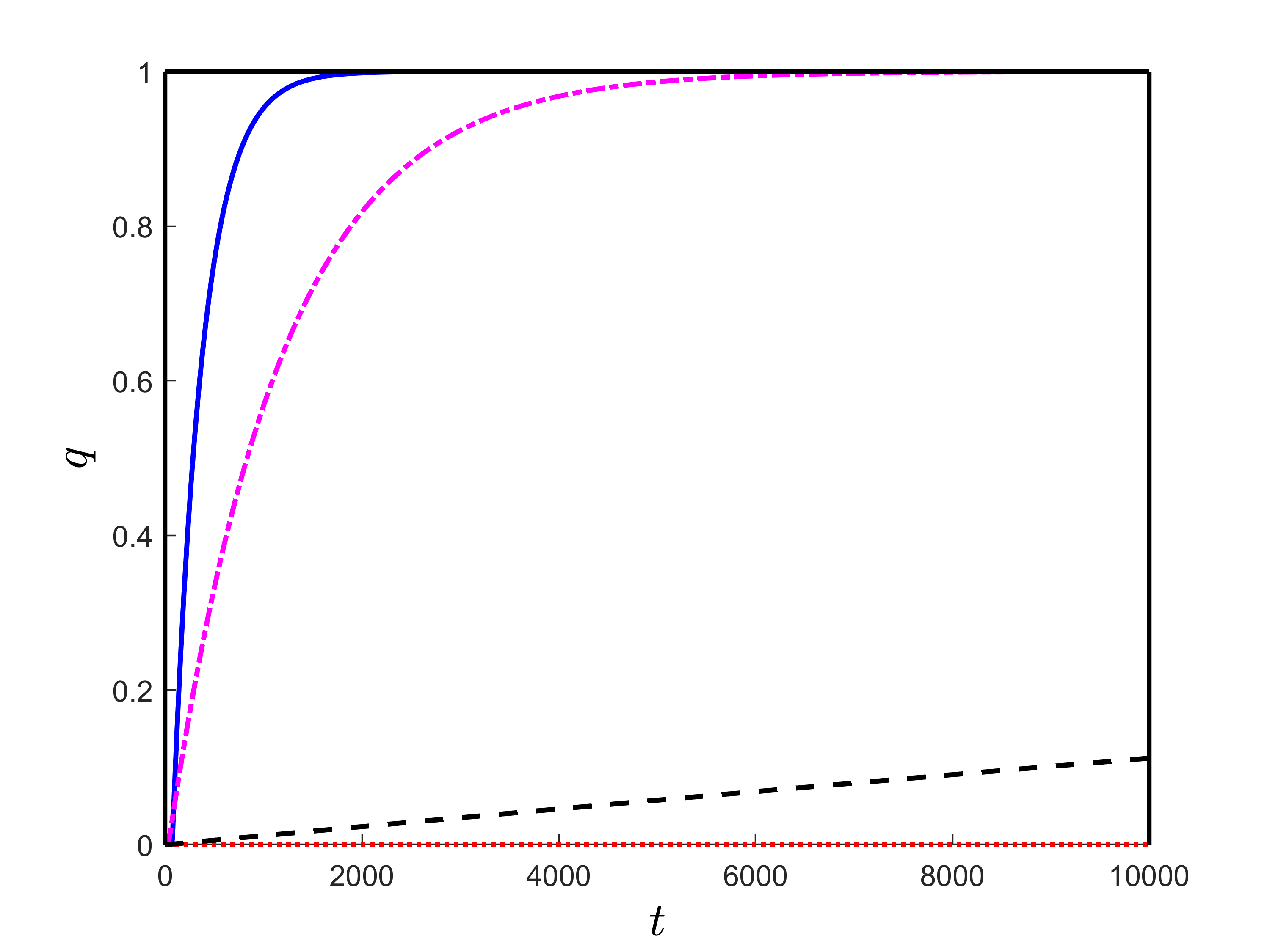}
                \caption{ }\label{fig:RWDSb}
        \end{subfigure}
        \caption{(Color online) Node-wise damage dynamics over time for local and nonlocal interactions in the neuronal damage model: (\subref{fig:RWDSa}) local and (\subref{fig:RWDSb}) nonlocal with $\sigma = 0.25$.}\label{fig:RWDS}
\end{center}
\end{figure}

The solutions of toxic $A\beta$ and toxic $\tau P$ converge to three different concentrations as demonstrated by Fig. \ref{fig:rwvbs}. But, the system has only one stable stationary point. This means the solution converges to some other stationary point(s) $E_{i} (0\leq i \leq 7)$. If a node is connected with other nodes, then the solutions converge to the positive stationary point $E_{*}$. On the other hand, suppose a node is not connected to the other nodes. Now, if this node is inside the seeding sites of toxic $A\beta$ or toxic $\tau P$, then the solution converges to $E_{6}$ or $E_{7}$, respectively; otherwise, it remains in ``healthy $A\beta$ - healthy $\tau P$'' state, i.e., converges to $E_{3}$. This happens due to the existence of stable manifolds of the stationary points $E_{3}$, $E_{6}$, and $E_{7}$, called semi-stable stationary points. These results are distinct from those obtained earlier \cite{TPEA1008267,SR2021,SR2021CP}. Indeed, if we consider the exponential growth in $A\beta$ and $\tau P$, then we can not get semi-stable stationary points \cite{TPEA1008267}.

\begin{figure}[]
\begin{center}
                \centering
                \includegraphics[width=10cm]{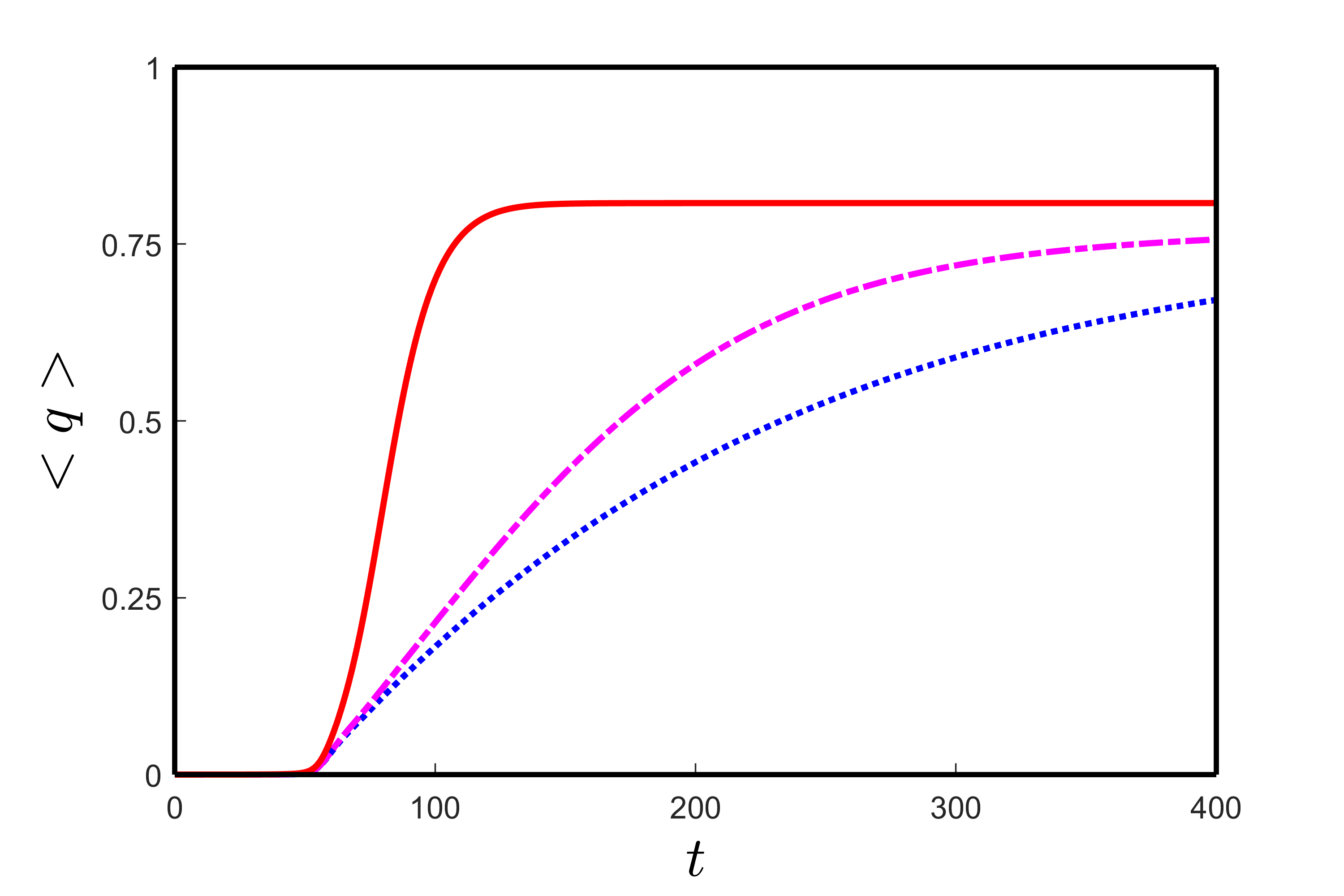}
        \caption{ (Color online) Spatial average of the damage of all the nodes with respect to time for different values of $k_{4}$: $k_{4}= 0.001$ (\hwplotka \hwplotka \hwplotka \hwplotka), $k_{4}= 0.01$ (\hwplotkb \hwplotkc \hwplotkb) and $k_{4}= 0.1$ (\hwplotkd).}\label{fig:LQSAK4}
\end{center}
\end{figure}

Now, we are in a position to analyze the neuronal damage of the nodes in the brain connectome in further detail. We solve the local and nonlocal versions of the neuronal damage model (\ref{NDM}) along with the local interaction of the model (\ref{HCMN}). In this case, we take the temporal parameter values from Table \ref{table:TFPV} along with $b_{3} = 4.14$, $c_{u}=c_{v}=0.1$. We choose $k_{1}$ less than $k_{2}$ to see the influence of toxic $\tau P$ neurofibrillary tangles on neural damage \cite{HC2016,CRJ2018,SLD2018} and $k_{3}$ larger than $k_{2}$, for enhancing the toxic effect of $\tau P$ in the presence of toxic $A\beta$ \cite{SLD2018,RE2017,SA2008,AL2011,EG2014}. Following these parameter restrictions, we fix $k_{1}=10^{-4}$, $k_{2} = 10^{-1}$ and $k_{3} = 10^{-2}$. For $k_{4} = 10^{-3}$, we plot the neuronal damage solutions for the local and nonlocal models ($\sigma = 0.25$) in Fig. \ref{fig:RWDS}. Similar to Fig. \ref{fig:rwvbs}, four types of damage profile exist for the neuronal damage model. Solid, dashed-dotted, dashed-dashed and dotted curves are corresponding to the neuronal damages of the nodes which are converging to the stationary points $E_{*}$, $E_{7}$, $E_{6}$ and $E_{3}$, respectively. The time taking to damage the nodes for the nonlocal model is larger compared to that of the local model. Also, an increase in the parameter $k_{4}$ causes faster neuronal damage in the brain connectome (see Fig. \ref{fig:LQSAK4}). 

For the case of secondary tauopathy, the stationary points $E_{5}$ and $E_{7}$ do not exist. Therefore, if a node is connected with other nodes or it is located in the seeding sites of toxic $\tau P$, solutions converge to the positive stationary point $E_{*}$. Now, if a node is disconnected and it is located inside the seeding sites of toxic $A\beta$, then the solution converges to $E_{6}$; otherwise, it remains in a healthy stationary state. In this case $E_{3}$ and $E_{6}$ are semi-stable stationary points.

The toxic propagations and the concentrations for the local and nonlocal network models are the same, corresponding to both the tauopathies. This happens due to the uniform parameter values all over the regions in the brain connectome. However, a qualitative change can be seen in the nonlocal interactions for non-uniform parameter values in the brain connectome. We study this behaviour in the following subsection.

\subsection{Mixed tauopathy}

We have studied the local and nonlocal network models with global constant parameter values. In the analysis of the whole living brain, this consideration of uniform parameters would need to be generalized further. The positron emission tomography (PET) imaging studies of $A\beta$ and $\tau P$ radiotracer uptake provide us with a better insight \cite{R2018}. In AD, the distribution of PET-$\tau$ SUVR intensities is biased in different parts of the brain. We incorporate this idea in the network model along with the AD patient data. We consider a sample parameter data provided in the Alzheimer’s Disease Neuroimaging Initiative (ADNI) database (adni.loni.usc.edu) \cite{TPEA1008267}.

\begin{table}[h!]
\caption{General synthetic parameter values}
\label{table:GSPV}
\begin{center}
\begin{tabular}{|c|c|c|c|}
\hline
Healthy $A\beta$ & Toxic $A\beta$ & Healthy $\tau P$ & Toxic $\tau P$ \\
\hline
$\rho = 1.38$ & $ \rho = 0.138$ & $\rho= 1.38$ & $\rho= 0.014$  \\
\hline
$a_{0} = 1.035$ & $\widetilde{a}_{1} = 0.828$ & $b_{0}= 0.69$ & $\widetilde{b}_{1}= 0.552$  \\
\hline
$a_{1} = 1.38$ & $a_{2} = 1.38$ & $b_{1}= 1.38$ & $b_{2}= 1.035$\\
\hline
\end{tabular}
\end{center}
\end{table}

\begin{table}[h!]
\caption{Modified $b_{3}$ values}
\label{table:MBV}
\begin{center}
\begin{tabular}{|c|c||c|c|}
\hline
 \multicolumn{4}{|c|}{Brain region ID and modified $b_{3}$ value} \\
\hline
Pars Opercularis & 7.452 & Rostral middle frontal gyrus & 6.707 \\
\hline
Superior frontal gyrus & 7.452 & Caudal middle frontal gyrus & 7.452 \\
\hline
Precentral gyrus & 5.589 & Postcentral gyrus & 3.726 \\
\hline
Lateral orbitofrontal cortex & 6.486 & Medial orbitofrontal cortex & 6.486 \\
\hline
Pars triangularis & 5.520e-6 & Rostral anterior cingulate & 6.210e-6 \\
\hline
Posterior cingulate cortex & 3.45 & Inferior temporal cortex & 13.11 \\
\hline
Middle temporal gyrus & 11.04 & Superior temporal sulcus & 8.97 \\
\hline
Superior temporal gyrus & 8.28 & Superior parietal lobule & 12.42 \\
\hline
Cuneus & 13.8 & Pericalcarine cortex & 13.8 \\
\hline
Inferior parietal lobule & 11.73 & Lateral occipital sulcus & 15.18 \\
\hline
Lingual gyrus & 13.8 & Fusiform gyrus & 7.59 \\
\hline
Parahippocampal gyrus & 11.04 & Temporal pole & 1.104e-5 \\
\hline
\end{tabular}
\end{center}
\end{table}

For the mixed tauopathy, we study a mixed version of primary and secondary tauopathies, i.e., some parts in the brain connectome satisfy secondary tauopathy, and the rest satisfies primary tauopathy. A mixed neuropathology with amyloid-beta plaques and tau pathology is present in nearly $40\%$ of patients and are associated with an increased rate of dementia and a decreased survival time \cite{DJ2017}. First, we set all the nodes in the brain connectome to a secondary tauopathy by considering the parameters in Table \ref{table:GSPV}, $b_{3} = 4.14$ and $c_{u}=c_{v} = 0.1$. According to the ADNI data, a modified value of the coupling parameter $b_{3}$ in some of the regions, given in Table \ref{table:MBV}, and the rest regions $b_{3} = 4.14$. The modification is symmetric, i.e., $b_{3}$ has the same value for both the left and right hemispheres in the corresponding regions. Motivated by \cite{TPEA1008267}, we also modify $b_{2}$ and $b_{3}$ in some of the regions (see Table \ref{table:BRBB}), so that the system has a state of primary tauopathy.

\begin{table}[h!]
\caption{Modified $b_{2}$ and $b_{3}$ values}
\label{table:BRBB}
\begin{center}
\begin{tabular}{|c|c|c|c|c|c|}
\hline
Brain region & Entorhinal cortex & Pallidum & Locus coeruleus & Putamen & Precuneus \\
\hline
$b_{2}$ & $ 3.125$ & $2.76$ & $1.38$ & $3.795$ & $3.105$\\
\hline
$b_{3}$ & 1.104e-5 & $2.76$ & $1.38$ &  $3.795$ & $3.105$\\
\hline
\end{tabular}
\end{center}
\end{table}

\begin{figure}[]
\begin{center}
                \centering
                \includegraphics[width=17cm]{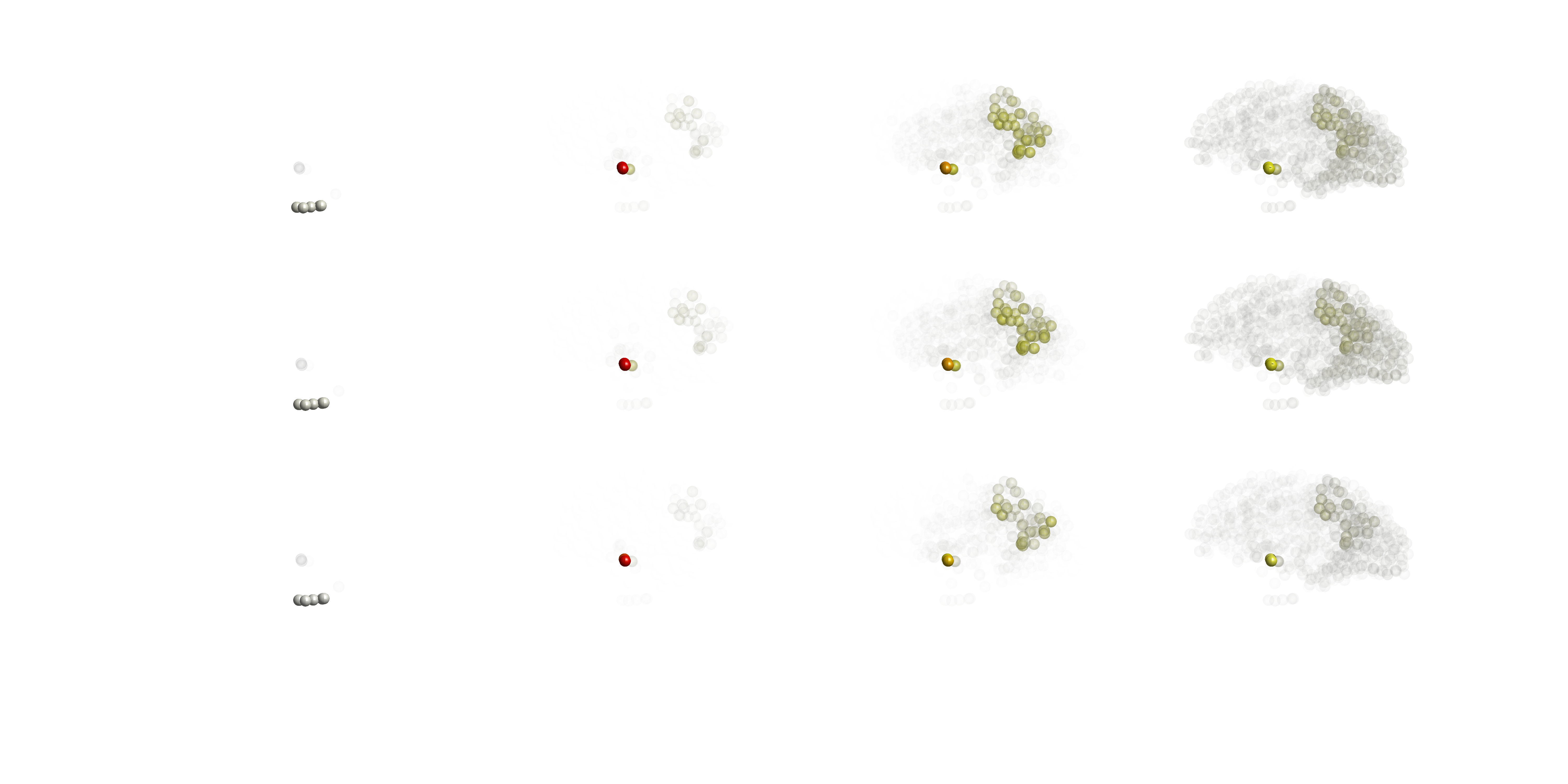}
        \caption{(Color online) Toxic-$\tau P$ distributions in the brain over time (left to right: $t=10, 20, 50$ and $100$) for the mixed tauopathy. Upper panel is corresponding to the local model, while middle and lower panels are corresponding to the nonlocal model with $\eta=0.5$ and $\eta=0.25$, respectively.}\label{fig:lnlvbmt}
\end{center}
\end{figure}

We use the same seeding sites as the initial condition for the mixed tauopathy. The solutions of the local and nonlocal models corresponding to the toxic $\tau P$ are shown in Fig. \ref{fig:lnlvbmt}. Due to the mixed parameter values in the brain region, the concentrations (all four components) are different in each node in the brain connectome. With decreasing the parameter values of $\eta$, we see a higher toxic density accumulation in some nodes. The targeted nodes depend on the degree of the node, i.e., the number of nodes connected with the node. Therefore, the spreading pattern of the toxic concentrations for the mixed tauopathy is different from primary or secondary tauopathies.

\begin{figure}[ht!]
\begin{center}
                \centering
                \begin{subfigure}[p]{0.4\textwidth}
                \centering
                \includegraphics[width=\textwidth]{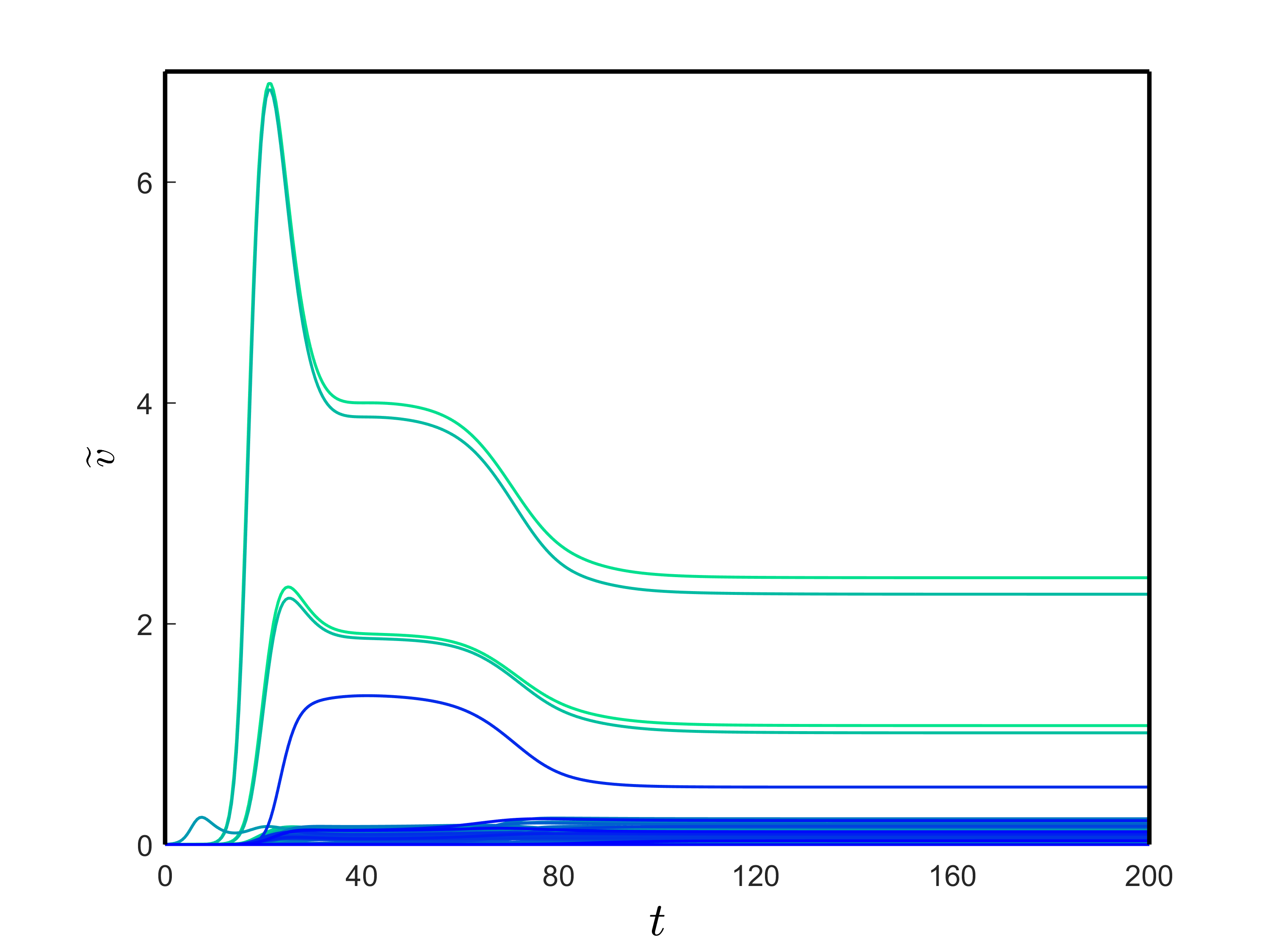}
                \caption{ }\label{fig:RWvbma}
        \end{subfigure}
        ~
        \begin{subfigure}[p]{0.4\textwidth}
                \centering
                \includegraphics[width=\textwidth]{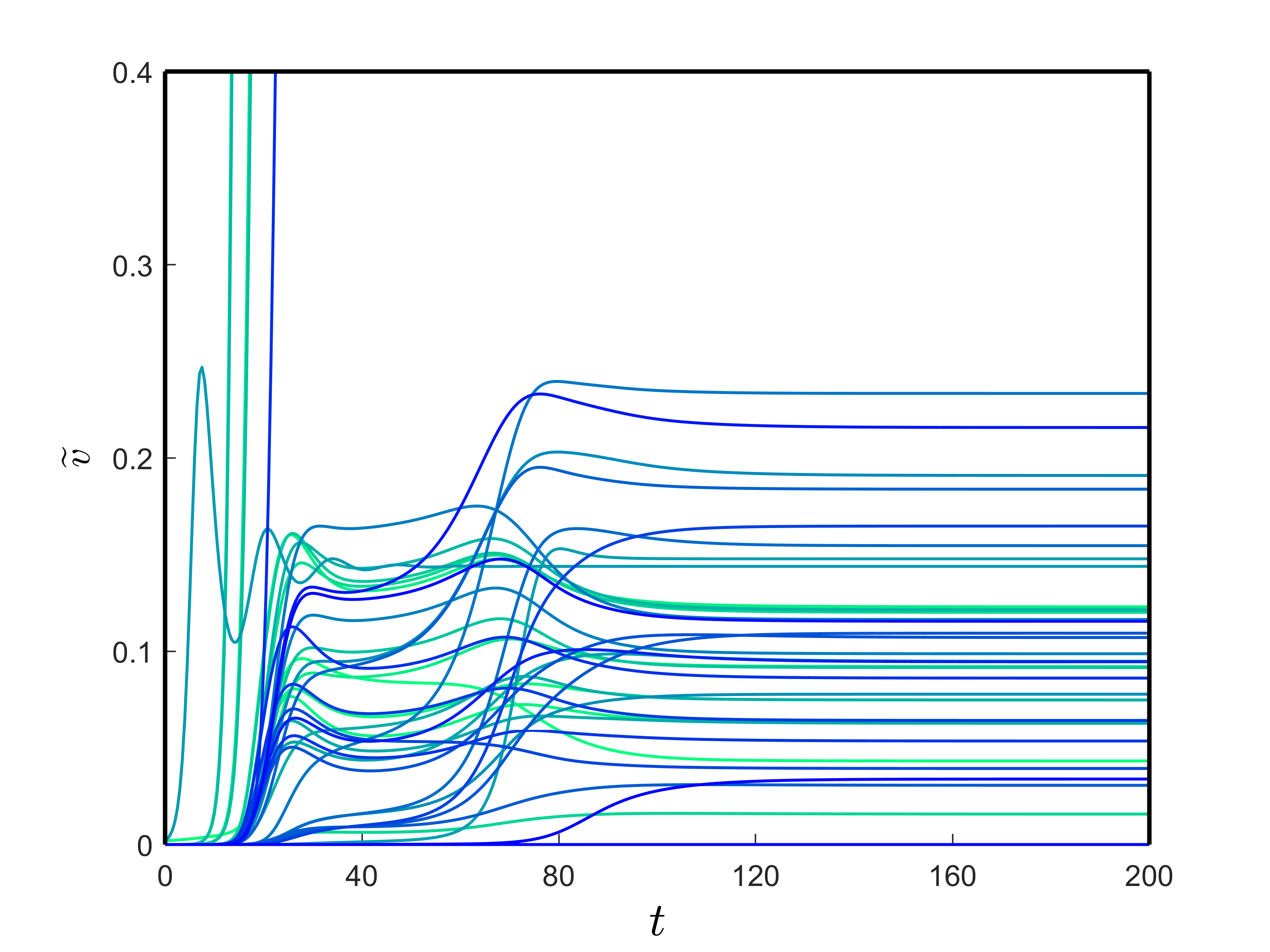}
                \caption{ }\label{fig:RWvbmb}
        \end{subfigure}
        \caption{(Color online) Node-wise toxic-$\tau P$ distributions over time for the nonlocal model ($\eta = 0.5$) with mixed tauopathy: (\subref{fig:RWvbma}) different curves corresponding to different brain regions; (\subref{fig:RWvbmb}) zoom version of (\subref{fig:RWvbma}).}\label{fig:RWvbm}
\end{center}
\end{figure}

\begin{figure}[ht]
\begin{center}
                \centering
                \includegraphics[width=8cm]{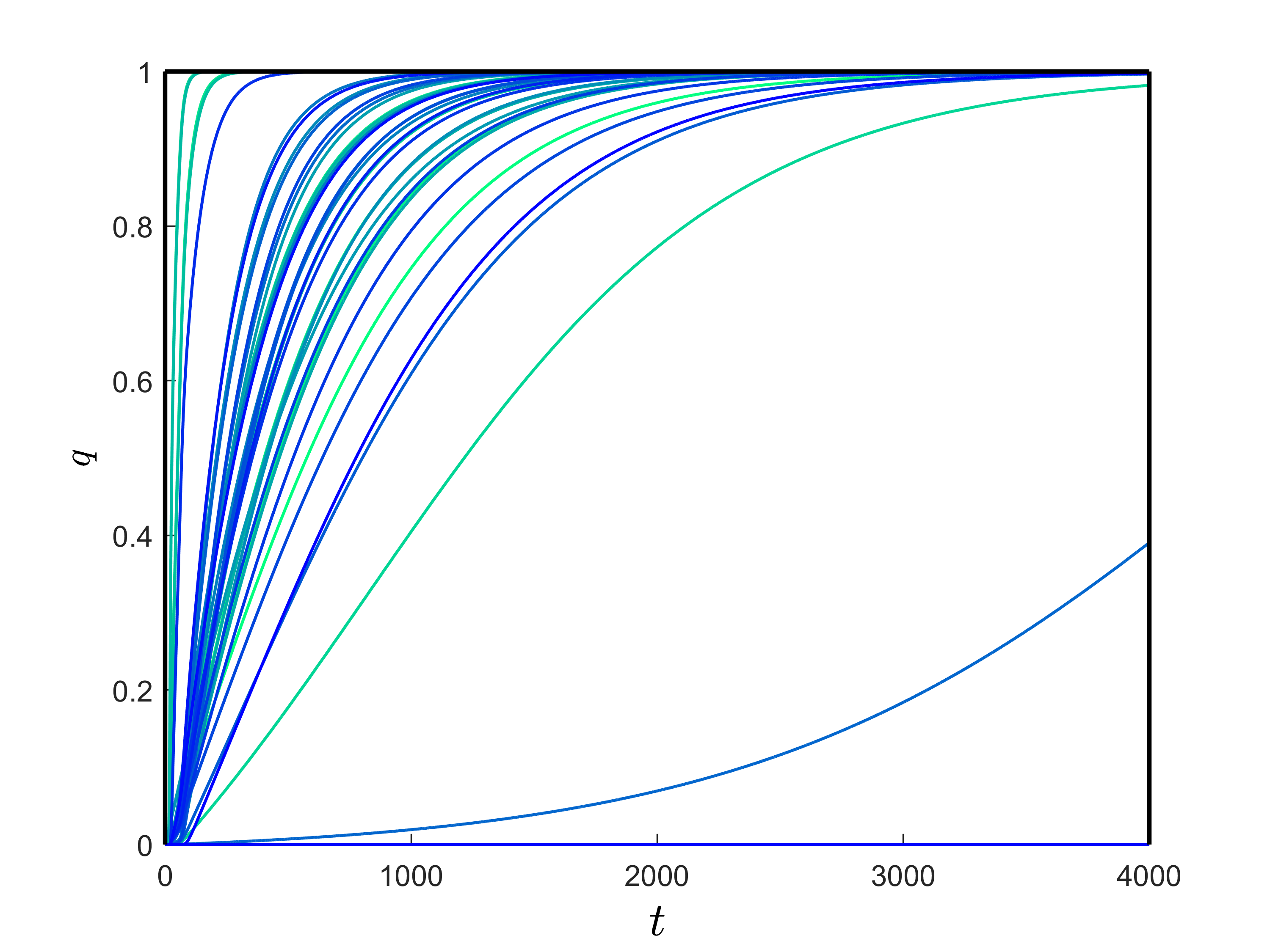}
        \caption{(Color online) Region-wise damage for the nonlocal model with mixed tauopathy.}\label{fig:nlmrwd}
\end{center}
\end{figure}

We plot the dynamics of toxic $\tau P$ in each of the regions in Fig. \ref{fig:RWvbm}. The spreading profiles of the toxic concentration in different nodes are different. Therefore, the time required to damage the nodes is different (see Fig. \ref{fig:nlmrwd}).

\section{Summary}{\label{SE5}}

A modified nonlocal coupled heterodimer multiscale model has been developed and applied for a better understanding of the dynamics of Alzheimer's disease. We have derived a nonlocal network model corresponding to the model based on a coupled system of integro-differential equations and incorporated the brain connectome data into this new model. Further, we have considered a nonlocal interaction in the damage dynamics. The stability behaviour of each of the stationary points of the system has been analyzed by using linear stability analysis. Two types of tauopathies (primary and secondary) have been discussed in detail, and their combination has been examined. 

For the primary tauopathy, the spreading patterns of the toxic concentration of $A\beta$ and $\tau P$ in the brain connectome are different in different regions. This occurs due to the initial seeding sites in the brain connectome. In this case, a total of four different spreading patterns are possible. On the other hand, three different types of spreading patterns exist for the secondary tauopathy. Our analysis has also revealed that many types of spreading patterns exist in the case of mixed tauopathy. The damage dynamics follow the same number of spreading patterns. Irrespective of the tauopathies, the nonlocal model takes longer to propagate the disease than the local model.

Given that tau is an MT-building protein, its more refined description can be achieved by including corresponding microtubules into consideration. Earlier developed models for the analysis of the microtubules \cite{SR2020a,SR2020b,SJ2020} can prove to be useful for further refinement of the models presented here. Further, it is now known that AD may affect not only the brain but the entire body, including the cardiovascular system \cite{JA2005}. Amyloid-beta protein fragments that form plaques in the brain of AD patients might stiffen their heart muscles due to the deposition of sticky amyloid-beta protein. Researchers discovered that Alzheimer's is caused by amyloid-beta proteins building up in the spaces between brain cells. While this causes noticeable symptoms in the brain, this same protein plaque can build up around the heart. The advancement of nonlocal models in this direction represents an important avenue of future research, where the methodology developed here may prove to be advantageous.

\bibliography{sample}

\section*{Acknowledgements}

The authors are grateful to the NSERC and the CRC Program for their support and RM is also acknowledging the support of the BERC 2018-2021 program and Spanish Ministry of Science, Innovation, and Universities through the Agencia Estatal de Investigacion (AEI) BCAM Severo Ochoa excellence accreditation SEV-2017-0718, and the Basque Government fund ”AI in BCAM EXP. 2019/00432”

\section*{Author contributions statement}


Both the authors contributed equally to this work.

\section*{Additional information}
 \textbf{Competing interests}
 Authors declare no conflict of interest.
 
 \noindent\textbf{Data availability}
 The data used in this article is available at www.braingraph.org.


\end{document}